\documentclass[aps,prl,twocolumn,showpacs,superscriptaddress]{revtex4}
\usepackage{graphicx}

\begin{document}

\title{Anisotropic softening of collective charge modes in the vicinity of critical doping in a doped Mott insulator}

\author{Y.W. Li}
\author{D. Qian}
\author{L. Wray}
\author{D. Hsieh}
\affiliation{Department of Physics, Joseph Henry Laboratories of
Physics, Princeton University, Princeton, NJ 08544}
\author{Y. Kaga}
\author{T. Sasagawa}
\author{H. Takagi}
\affiliation{Department of Physics, University of Tokyo, Tokyo
113-8656, Japan}
\author{R.S. Markiewicz}
\author{A. Bansil}
\affiliation{Department of Physics, Northeastern University, Boston,
MA 02115}
\author{H. Eisaki}
\affiliation{AIST, 1-1-1 Central 2, Umezono, Tsukuba, Ibaraki,
305-8568 Japan}
\author{S. Uchida}
\affiliation{Department of Physics, University of Tokyo, Tokyo
113-8656, Japan}
\author{M.Z. Hasan}
\affiliation{Department of Physics, Joseph Henry Laboratories of
Physics, Princeton University, Princeton, NJ 08544}

\begin{abstract}

Momentum resolved inelastic resonant x-ray scattering is used to map
the evolution of charge excitations over a large range of energies,
momenta and doping levels in the electron doped Mott insulator class
Nd$_{2-x}$Ce$_x$CuO$_4$. As the doping induced AFM-SC
(antiferromagnetic$-$superconducting) transition is approached, we
observe an anisotropic softening of collective charge modes over a
large energy scale along the $\Gamma\rightarrow
(\pi,\pi)$-direction, whereas the modes exhibit broadening ($\sim$ 1
eV) with relatively little softening along $\Gamma\rightarrow
(\pi,0)$ with respect to the parent Mott insulator (x=0). Our study
indicates a systematic collapse of the gap consistent with the
scenario that the system dopes uniformly with electrons even though
the softening of the modes involves an unusually large energy scale.

\end{abstract}


\pacs{78.70.Ck, 71.20.-b, 74.25.Jb}

\date{\today}

\maketitle


The evolution of a strongly correlated material with doping$-$from a
Mott insulator to a conducting metal$-$is one of the most
intensively studied issues in current condensed matter physics. This
fascinating evolution has proven to be full of surprises such as the
appearance of high-T$_c$ superconductivity, non-Fermi liquid
behavior, and nanoscale phase separation \cite{imada}. Mott
insulators often exhibit phase transitions upon doping, which are
signaled or hallmarked by the softening of collective charge or spin
modes. The behavior of spin modes has been investigated extensively
via neutron scattering \cite{rb}. Although charge excitations near
the Brillouin zone (BZ) center can be accessed by optical techniques
\cite{klein}, their behavior with momentum over the full BZ remains
largely unexplored. Here, as demonstrated in recent experimental
\cite{kim/pa/hill, hasan/isaacs} and theoretical studies
\cite{PP/TT/TD, MarB}, inelastic x-ray scattering indeed provides
such a unique opportunity.

While previous studies have focused largely on either the undoped
1-D \cite{1d} or 2-D \cite{kim/pa/hill, hasan/isaacs} insulators, or
the hole-doped superconductors \cite{kim/ishii/lu}, Mott insulators
can be doped with electrons as well. In fact, it appears that with
electron doping bands in the cuprates evolve in a much more
straightforward and systematic manner
\cite{KLBM,KuR,SeMSTr,BKAT,YYT} than with hole doping. However, a
direct study of the doping induced changes in the collective
excitation spectra of a Mott insulator is lacking. Previous x-ray
scattering work \cite{ITE} focused on the highly electron doped
\textit{superconductor} (SC) Nd$_{1.85}$Ce$_{x=0.15}$CuO$_4$,
observed intraband and interband excitations and interpreted the
excitations in a one-band model with long-range hoppings. However,
Ref[15] does not report the undoped Mott insulator and the way gap
modes of the Mott insulator are related to the doping where
superconductivity sets in. In this Letter, we report a
high-resolution study of how the collective charge excitations of
the Mott insulating state ($x$=0) develop with electron doping in
approaching the critical AFM-SC transition (before reaching the
optimal doping) for the first time. Our finding, made possible by
studying the insulating states, is that, as the electron-doping
induced AFM-SC transition is approached from the x=0 Mott side, the
system exhibits an anisotropic softening of excitations over a large
energy scale ($\sim$ 5 eV) along the $\Gamma\rightarrow
(\pi,\pi)$-direction, whereas the modes exhibit broadening ($\sim$ 1
eV) with relatively little softening along $\Gamma\rightarrow
(\pi,0)$. Our results suggest that a multi-orbital Hubbard model is
essential to describe the observed evolution from the x=0 Mott state
in contrast to the 1-band model with long-range hoppings proposed in
Ref[15] based on the superconductor only. Moreover, our results show
that the \textit{evolution} of Mott physics of the electron-doped
cuprate is dramatically different from that reported in the
hole-doped cuprates \cite{kim/ishii/lu}.

\begin{figure}[t]
\includegraphics[width=8cm]{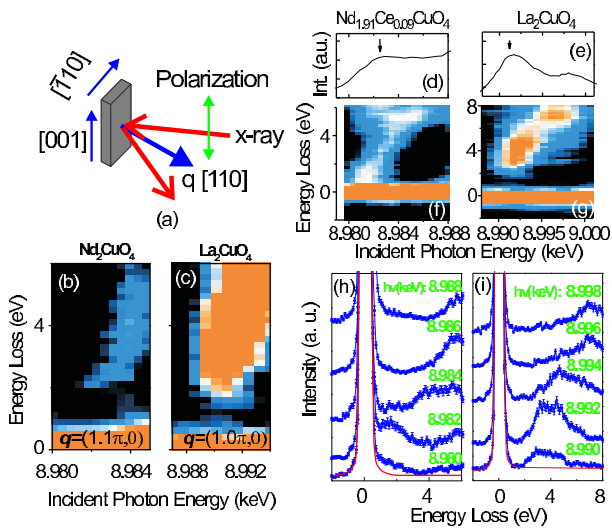}
\caption{Incident energy dependence of electronic excitations: (a)
Vertical scattering geometry employed with x-ray field along [001].
(b) and (c) X-ray energy dependence of inelastic excitations for
Nd$_2$CuO$_4$ and La$_2$CuO$_4$. Q-values are chosen to minimize
large quasielastic background.
(d)-(g) Excitations near the Mott
gap are seen to be enhanced near the first absorption peak, not only
in the undoped insulator, but also in the doped system. A few
energy-loss curves corresponding to the data images in (f) and (g)
are shown in (h) and (i). The red curves show the fits to the
quasielastic data.}
\end{figure}

The electronic structure of Nd$_{2-x}$Ce$_x$CuO$_4$ (NCCO) has been
studied by angle-resolved photoemission (ARPES) and optical
spectroscopies. ARPES studies \cite{nparm} find that the electrons
dope directly into the bottom of the upper Hubbard band and yield
small Fermi (FS) pockets, with a crossover around optimal doping to
a large FS. Such a scenario, where the magnetic order remains
commensurate without signs of `stripe' (charge inhomogeneity) or
other phase separation, also describes
magnetization\cite{Gre,MKII,BKAT} and optical data\cite{Mill}.

The experiments were performed at CMC-CAT beamline 9-ID-B at the
Advanced Photon Source. Resonant inelastic x-ray scattering (RIXS)
at copper $K$ edge allows a large enough momentum transfer to cover
several BZ's. The scattered photon energy was measured by using a
diced Ge (733) crystal analyzer, and the intensity was recorded by a
solid state detector. The overall energy resolution was set to about
0.37 eV in order to \textit{improve count} efficiency, enabling us
to detect fine momentum-dependent details of the spectra. All data
are taken at room temperature. To avoid possible polarization
induced artifacts when measuring the in-plane anisotropy, the sample
was mounted with incident polarization directed along the $c$-axis
(Fig.1(a)). The data were collected at several values of the
momentum transfer vector $\mathbf{q=k_i-k_f}$ in the 2nd BZ along
the [100] direction, and in the 4th BZ along the [110] direction.
The coupling to various excitations in the x-ray scattering process,
near a resonance, depends in general on the energy of the incident
photons. Accordingly, we first examined the detailed incident energy
dependence of the loss spectra at several momenta. Representative
data sets are presented in Fig.~1. It is known that the
charge-transfer gap excitations resonate near the first absorption
peak \cite{kim/ishii/lu}. Similar resonance behavior is seen in
Nd$_2$CuO$_4$, which is like that in the more extensively studied
La$_{2-x}$Sr$_x$CuO$_4$ \cite{kim/ishii/lu}. However, the scattering
intensity at the low-energy branch is about an order of magnitude
\textit{weaker}. Our systematic investigations summarized in Fig. 1
show that similar shape of the resonance profile is seen for the
lower energy excitations in the electron doped system. Accordingly,
we employed the photon energy of 8.982 keV (this energy is above the
1s$-$$>$3d edge which suppresses the crystal-field excitations
\cite{seo} as it is not the focus of the current study). Therefore,
in the theoretical simulations such incident energy correspondences
of the excitation spectra are maintained for comparison with the
measured data.

Fig.2 summarizes our RIXS data on NCCO. In the undoped system in
panel (a), a broad excitation is observed near the zone center
around 2 eV with an onset energy of $\approx$ 1 eV, which is
consistent with the charge transfer gap found in this compound in
optical studies \cite{tokura}. It is seen to lose intensity as it
approaches the zone corner along the [$\pi$,$\pi$] as well as the
[$\pi$,0] directions (Fig.2(a)). Along [$\pi$,$\pi$], it merges at
the zone corner with a higher excitation band near 5 eV which
disperses upward. With doping these two high energy branches split
at $(\pi,\pi)$ as seen in (b) and (c), but only the lower energy
branch softens significantly (i.e. moves to lower energy), so that
the overdoped system in (c) displays a large excitation gap at the
$(\pi,\pi)$ point from 2 eV to 4 eV. In sharp contrast, the
aforementioned 2 eV branch evolves uniformly along [$\pi$,0], and
softens rapidly near the zone center as it tends to close the
excitation gap with doping. Notably, the zero-loss energy is not
accessible due to the presence of the strong quasielastic peak (see,
e.g., Fig.1(h)), making it difficult to extract significant data
below $\approx$ 0.4 eV, where fitting and subtraction of the
quasielastic peak lead to uncertainties. Nevertheless, the softening
of the low energy excitations is evident in the data over the BZ.

The preceding observations are further highlighted through the
directly measured data curves presented in Fig.3. Doping dependent
changes in the individual energy-loss curves are compared in (a).
Along [110], doping splits the peak in the undoped spectrum around 5
eV at $(\pi,\pi)$ (red uppermost curve on the left side of (a)) into
two peaks, and the lower of these peaks softens rapidly with doping.
In contrast, along [100] a monotonic softening of the low energy
excitations is found. Changes in the positions of various spectral
features and some of the associated leading edges are plotted in (b)
as a function of momentum and doping. The upper branches around 5-6
eV are seen to be affected weakly by doping. Despite a significant
softening of one of the modes around $(\pi,\pi)$, the closing of the
excitation gap is limited to the region of the zone center as one
approaches the superconducting phase ($x=0.14$). These results are
in clear contrast to the behavior in hole-doped cuprates such as the
La$_{2-x}$Sr$_x$CuO$_4$ or the $Y$Ba$_2$Cu$_3$O$_{7+d}$
series\cite{kim/ishii/lu}.

\begin{figure}[t]
\includegraphics[width=9cm]{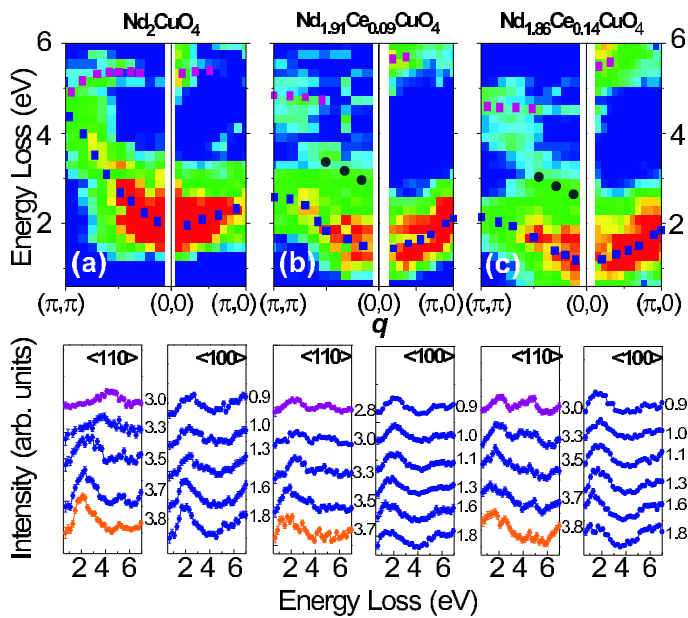}
\caption{ Doping evolution of charge excitations in NCCO. (a) $x$=0
(parent Mott state), (b) $x$=0.09 (doped AFM), and (c) $x$=0.14
(superconductor). Upper row shows intensity maps in the reduced-zone
using a color scheme where high intensity is denoted by red and low
by blue.  Black, grey and red dots are guides to the eye for the
dispersion of low, medium and high energy excitation branches,
respectively. Lower row gives a few representative spectra
corresponding to the images in the upper row. Absolute momentum
values (in units of $\pi$/a$_o$) are shown attached to various
spectra and lie in the 4th BZ along [110] and the 2nd BZ along
[100]. }
\end{figure}

\begin{figure}[t]
\includegraphics[width=8cm]{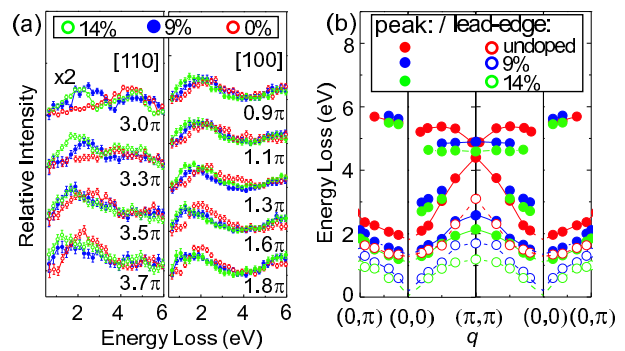}
\caption{ (a) Spectral curves (raw data) along [110] and [100] for
the undoped system (red points) are compared at various momenta with
the corresponding spectra for the doped system (black and green
points). (b) Positions of peaks (marked in Fig.-2(a-c)) based on the
center of gravity method \cite{hasan/isaacs, 1d} of various spectral
features (filled circles) and some of the associated leading edges
(open circles) as a function of momentum and doping are plotted. }
\end{figure}

Recent theoretical analyses \cite{MarB, ITE} of data on x=0.15
samples suggest two different scenarios. The analysis of the data
using a 1-band model Ref[15] suggests a no gap collapse scenario
whereas the analysis of high doping data in Ref.[7] suggests a gap
collapse scenario. Which of these two opposite viewpoints is correct
cannot be ascertained from the data of Ref.[15] alone. Therefore it
is crucial to study the evolution of the undoped compound of this
series with high resolution so that the low-energy gap excitations
can be accessed and model the evolution of the spectra from the x=0
Mott side. To interpret the present data in terms of RIXS spectra
computed within the framework of a three-band Hubbard Hamiltonian of
NCCO, based on Cu $d_{x^2-y^2}$ and two O $p_{\sigma}$ orbitals, we
extend the theoretical framework described in \cite{MarB} by
incorporating the doping evolution data made available here for the
first time. The specific values of the parameters used in this work
to fit the spectra are : $t_{CuO}=0.85$ eV, $t_{OO}=-0.6$ eV,
$\Delta_0 =-0.3$ eV, $U_p=5.0$ eV; where $t_{CuO}$ and $t_{OO}$ are
the Cu-O and O-O nearest neighbor (NN) hopping parameters, $n_d$
[$n_p$] is the average electron density on Cu [O], $m_d$ the average
electron magnetization on Cu, and $U$ [$U_p$] is the Cu [O] on-site
Coulomb repulsion. The remaining parameter, crucial for
\textit{fitting the doping evolution of the excitations}, the
Hubbard $U$, is taken to be $7.45$ eV at $x$=0 with weak doping
dependence: $U$=6.69 eV at $x$=0.09 and $U$=6.27eV at $x$=0.14, so
that the effective $U$ decreases by about 16\% over this doping
range, reflecting presumably the effects of screening. We note that
before proceeding with RIXS computations, we determined the chemical
potential and the magnetization $m_d$ on Cu sites self-consistently
at each doping level. $m_d$ values so found are: 0.32 at $x$=0; 0.19
at $x$=0.09; and 0.12 at $x$=0.14.

Figure~4 (top) shows the calculated RIXS intensity maps. The
positions of various experimentally observed spectral peaks (filled
circles) and the leading edge of the low energy feature (filled
diamonds) are superposed for ease of comparison. The high energy
peaks involve transitions from the nonbonding O and bonding Cu-O
bands to unoccupied states in the antibonding band, and fall in the
same energy range as transitions involving other Cu and O orbitals
not included in the present 3-band model. Therefore, it is
appropriate to concentrate on the behavior of RIXS peaks within a
few electron volts, which are associated with the antibonding Cu-O
band, split by AFM ordering. In this energy region the theoretically
predicted changes in the energies of various features as a function
of momentum and doping are in reasonable accord with experimental
observations highlighted in the discussion of Figs.2 and 3 above. In
particular, the softening of low energy peaks is well reproduced and
the collapse of the gap occurs only near $\Gamma$ in the
computations.  The latter effect can be readily understood. The RIXS
transitions involve both intraband and interband contributions. In
the absence of a gap, only {\it intraband} transitions are possible
near $\Gamma$ which can only exist near zero energy transfer close
to the Fermi level.  Away from $\Gamma$, intraband transitions can
take place at finite energies, but when a gap opens up, {\it
interband} transitions become allowed, \textit{even at $\Gamma$}
\cite{4ev}.

Although the computed maps in the top row of Fig.4 are shown
unbroadened to highlight spectral features, the theoretical spectra
shown in the bottom panels of Fig.4 (blue lines) have been smoothed
to mimic the broadening of experimental lineshapes\cite{broaden}.
The relative intensities of the computed peaks are seen to be in
agreement with our experimental results in lower
energies\cite{highout}, excepting the momentum region near the $(\pi
,\pi )$-point, where both the experimental and theoretical
intensities are weak, but the computed cross-section is smaller. A
similar behavior is seen in the La$_{2-x}$Sr$_x$CuO$_4$ series
\cite{kim/ishii/lu}. Excitonic effects associated with longer range
Coulomb coupling (intersite-V) beyond that included in our model
computations could enhance and redistribute spectral weight near
zone boundaries \cite{SP}. The \textit{low-energy, gap-edge}
dispersion relations (E vs. q) are in agreement as seen from the top
row of Fig-4(a-c)\cite{4ev}.

\begin{figure}[t]
\includegraphics[width=8cm]{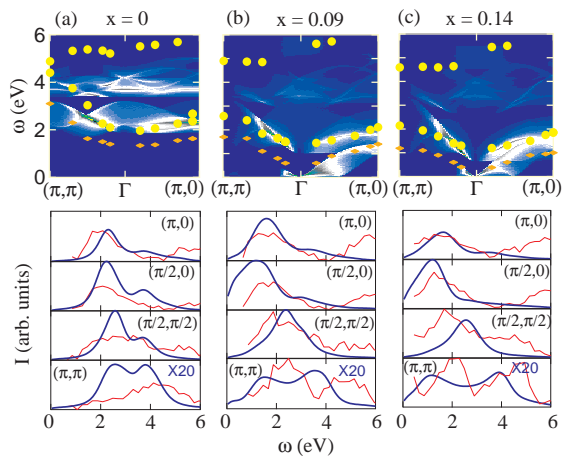}
\caption{{\it Top row}: Doping evolution of RIXS spectra obtained
within a three-band Hubbard model as a function of energy transfer
$\omega$ at momenta ${\bf q}$ along $\Gamma$ to $(\pi,\pi)$ and
$(\pi,0)$ directions for three different dopings $x$. Positions in
$\omega-{\bf q}$ space of various experimentally observed spectral
peaks (filled circles) and the leading edge of the low energy
feature (filled diamonds) are superposed. Computed spectra have not
been broadened to reflect experimental resolution in order to
highlight spectral features. {\it Bottom row}: Comparison of
theoretical (blue lines) and experimental (red lines) spectra. Three
columns refer to the three indicated doping levels $x$. Each column
includes four different $q$-values as shown; experimental spectra
are at the closest $q$-value measured. Theoretical spectra have been
smoothed to reflect experimental broadening\cite{broaden}. }
\end{figure}

\begin{figure}[t]
\includegraphics[width=9cm]{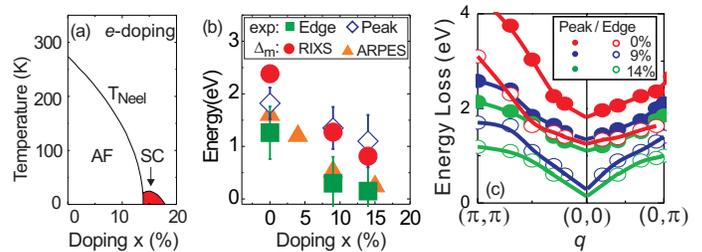}
\caption{ (a) Schematic temperature-doping ($T-x$) phase diagram of
NCCO. (b) The leading-edge gap (green squares) and peak positions
(white diamonds) of the lowest energy excitation branch in RIXS data
at the $\Gamma$-point are compared with AFM correlation gap
$\Delta_m$ from the present calculation (filled circles) and from
ARPES (triangles)\cite{KLBM} as a function of doping. (c) The
excitation modes (branches) are fitted and extrapolated to the
$\Gamma$-point(0,0) based on the branch curvatures, to extract the
data points presented in (b). Peak position are shown in the raw
data in Fig-2(a-c) and leading-edges are from data in Fig-3.}
\end{figure}


Taken together, these results suggest that the parent x=0 Mott
insulating state of NCCO is doped uniformly by the addition of
electrons (within the resolution) and that the AFM-gap collapses
near the onset of the SC phase as seen from the systematics in
Fig.5. In Fig.5(b) we plot the doping dependence of the leading-edge
gap (edge, green squares) and peak position (peak, white diamonds)
of the lowest branch at $\Gamma$. [Since quasielastic scattering
makes it difficult to obtain experimental data right at $\Gamma$,
these values are found by extrapolation from nearby points, as shown
in Fig.5(c).] Within the experimental uncertainties the gap
collapses in the vicinity of the $\Gamma$-point as one approaches
the critical doping regime in the phase diagram. For comparison, we
also show in Fig.5(b) $\Delta_m=Um_d$, the AFM-correlation
gap\cite{MarB,KLBM} based on our present calculation (filled
circles) and the corresponding ARPES\cite{KLBM,foot1} results
(triangles), which are seen to be in agreement. In general, our
analysis suggests that the anisotropic softening involving a large
energy scale observed near the Mott insulating state requires going
beyond the one band model.


In summary, we have utilized the unique momentum resolution of x-ray
scattering to map the evolution of particle-hole excitations in the
electron doped cuprate from its parent Mott state (x=0) to the
doping on-set for superconductivity. We observe a nearly degenerate
charge excitation mode near the $(\pi,\pi )$ point around 5 eV in
the Mott insulator, which splits on doping away from half-filling,
with its lower energy branch softening anisotropically in
approaching the AFM-SC critical doping. In contrast, the response
near the ($\pi$,0) wavevector exhibits damping with relatively
little softening. Our results indicate a systematic collapse of the
gap consistent with the scenario of \textit{uniform doping} when
electrons are added into the parent Mott state which is in clear
contrast to what is reported in \textit{hole-doped} Mott insulators
where charge inhomogeneity is commonly observed. The unusual
softening of the modes from the Mott state likely contributes to
many unusual behaviors of cuprates.

We acknowledge T. Gog and D. Casa for technical assistance. This
work is primarily supported by DOE/DE-FG02-05ER46200. Use of the
Adv. Photon Source was supported by DOE/W-31-109-Eng-38. The
theoretical part of the work is supported by the US Department of
Energy contracts DE-AC03-76SF00098 and DE-FG02-07ER46352, and
benefited from the allocation of supercomputer time at the NERSC and
Northeastern University's Advanced Scientific Computation Center
(ASCC).


\begin{thebibliography}{99}
\bibitem{broaden}The spectra were first smoothed with a
Gaussian of 0.37 eV width to account for the resolution. The spectra
were further smoothed via folding with a Lorentzian of 0.30 eV width
to mimic what appears to be an intrinsic linewidth.
\bibitem{4ev}The branch near 4 eV is weak in the experimental data.
Since this branch has a different symmetry than the 2 eV branch it
resonates at a different photon energy (near 8.894 eV) as evident
from the raw data profile in Fig-1(f).
\bibitem{highout} The experimental spectra for energy transfers beyond
about 6 eV contain significant contributions from bands beyond the
3-band model we considered for this work.
\bibitem{foot1}In the one-band model, $U$ has a different definition, and the
magnetic gap is $\Delta_m=2Um_d$.  Also, at $x$=0, ARPES does not
see the upper band giving a lower limit to $\Delta_m$.
\end{thebibliography}
\end{document}